\documentclass[sigconf,screen]{acmart}

\AtBeginDocument{%
  }

\setcopyright{acmlicensed}
\copyrightyear{2018}
\acmYear{2018}
\acmDOI{XXXXXXX.XXXXXXX}

\acmBooktitle{Companion Proceedings of the 33rd ACM Symposium on the Foundations of Software Engineering (FSE '25), June 23--27, 2025, Trondheim, Norway}
\acmISBN{978-1-4503-XXXX-X/18/06}

\usepackage{amsmath,amsfonts}
\usepackage{algorithmic}
\usepackage{graphicx}
\usepackage{textcomp}
\usepackage{listings}
\usepackage{xcolor}
\usepackage{hyperref}
\usepackage{url}

\definecolor{codegreen}{rgb}{0,0.6,0}
\definecolor{codegray}{rgb}{0.5,0.5,0.5}
\definecolor{codepurple}{rgb}{0.58,0,0.82}
\definecolor{backcolour}{rgb}{1,1,1}

\newboolean{showcomments}
\setboolean{showcomments}{true}         
\ifthenelse{\boolean{showcomments}}
  {\newcommand{\nb}[2]{
  \fbox{\bfseries\sffamily\scriptsize#1}
     {\sf\small$\blacktriangleright$\textit{\textcolor{red}{#2}}$\blacktriangleleft$}
   }
  }
  {\newcommand{\nb}[2]{}
   
  }

\lstdefinestyle{mystyle}{
    backgroundcolor=\color{backcolour},   
    commentstyle=\color{codegreen},
    keywordstyle=\color{magenta},
    numberstyle=\tiny\color{codegray},
    stringstyle=\color{codepurple},
    basicstyle=\ttfamily\footnotesize,
    breakatwhitespace=false,         
    breaklines=true,                 
    captionpos=b,                    
    keepspaces=true,                 
    numbers=left,                    
    numbersep=5pt,                  
    showspaces=false,                
    showstringspaces=false,
    showtabs=false,                  
    tabsize=2
}

\begin{document}

\title{PCLA: A Framework for Testing Autonomous Agents in the CARLA Simulator}

\author{Masoud Jamshidiyan Tehrani}
\email{masoud.jamshidiyantehrani@usi.ch}
\affiliation{%
  \institution{Università della Svizzera italiana}
  \city{Lugano}
  \country{Switzerland}
}

\author{Jinhan Kim}
\email{jinhan.kim@usi.ch}
\affiliation{%
  \institution{Università della Svizzera italiana}
  \city{Lugano}
  \country{Switzerland}
}

\author{Paolo Tonella}
\email{paolo.tonella@usi.ch}
\affiliation{%
  \institution{Università della Svizzera italiana}
  \city{Lugano}
  \country{Switzerland}
}

\renewcommand{\shortauthors}{Tehrani et al.}

\begin{abstract}
Recent research on testing autonomous driving agents has grown significantly, especially in simulation environments. The CARLA simulator is often the preferred choice, and the autonomous agents from the CARLA Leaderboard challenge are regarded as the best-performing agents within this environment. However, researchers who test these agents, rather than training their own ones from scratch, often face challenges in utilizing them within customized test environments and scenarios. To address these challenges, we introduce PCLA (Pretrained CARLA Leaderboard Agents), an open-source Python testing framework that includes nine high-performing pre-trained autonomous agents from the Leaderboard challenges. PCLA is the first infrastructure specifically designed for testing various autonomous agents in arbitrary CARLA environments/scenarios. PCLA provides a simple way to deploy Leaderboard agents onto a vehicle without relying on the Leaderboard codebase, it allows researchers to easily switch between agents without requiring modifications to CARLA versions or programming environments, and it is fully compatible with the latest version of CARLA while remaining independent of the Leaderboard’s specific CARLA version. PCLA is publicly accessible at \url{https://github.com/MasoudJTehrani/PCLA}.
\end{abstract}




\maketitle

\section{Introduction}

The autonomous vehicle industry is rapidly growing, with extensive research focused on testing and improving the robustness of autonomous driving agents (ADAs)~\cite{ma2024slowtrack,von2023deepmaneuver,wang2023does,yoon2023learning,zhang2023data}. Testing these agents in challenging scenarios is an active area of research that requires either access to a real-world autonomous vehicle in a controlled test track or a simulation that replicates realistic environments~\cite{tehrani2024taxonomy}. Since simulations are more cost-effective, they are the preferred solution for extensive test campaigns. Thus, they play a crucial role in autonomous vehicle testing. Among the available simulation platforms~\cite{li2024choose}, CARLA~\cite{Dosovitskiy17} stands out as the leading choice. It is open-source, supported by a large community, actively maintained, fully customizable, and it offers features such as multiple weather conditions and five different types of sensors~\cite{li2024choose}.

To further advance the development of robust ADAs within the CARLA platform, a competitive challenge, the CARLA Leaderboard~\cite{leaderboard}, was introduced. This challenge allows participants to test their agents against a series of problematic scenarios and compare their performance on the Leaderboard. This approach helps developers make the `Leaderboard agents' robust and possibly ready for real-world testing/deployment. To ensure fair competition, the CARLA Leaderboard provides predefined codes that all participants must use to create and test their agents within the Leaderboard’s specified scenarios. Participants are not permitted to modify these scenarios or operate their agents outside the Leaderboard’s ecosystem. These constraints make the agents entirely dependent on the Leaderboard's codebase, rendering them at the same time unusable without it. Researchers who wish to deploy these agents in custom test scenarios or environments, must first fully understand the Leaderboard's code and then modify it according to their needs.

To address this issue and simplify the use of the Leaderboard's ADAs on new test environments/scenarios, we introduce PCLA, the Pretrained CARLA Leaderboard Agents. This open-source framework allows the users to easily deploy their desired ADA onto a vehicle without relying on the Leaderboard's codebase. With PCLA, researchers can operate their vehicle using their selected ADA, within their own developed CARLA environment, and they are not forced to stick to a specific version of CARLA, as required by the Leaderboard. PCLA is compatible with the latest version of CARLA and simplifies the process of switching between different agents. It provides a uniform interface, eliminating the burden of adapting to different programming environments or CARLA versions for each agent. Moreover, instead of moving the vehicle, PCLA outputs the next vehicle movement parameters (next action) for the next frame, so that researchers have the flexibility to apply it to a vehicle for movement or use it in other applications (e.g., system level attack generation).
To the best of our knowledge, this is the first infrastructure designed for testing various autonomous agents in the CARLA simulator.

The six contributions of this work are as follows:
\begin{itemize}
    \item We introduce a new infrastructure named PCLA, for testing various autonomous agents in the CARLA simulator.
    \item PCLA provides a clear method to deploy ADAs onto a vehicle without relying on the Leaderboard codebase.
    \item PCLA enables easy switching between ADAs without requiring changes to CARLA versions or programming environments.
    \item PCLA provides the next movement action computed by the chosen agent, which can then be utilized in any desired application.
    \item PCLA is fully compatible with the latest version of CARLA, and independent of the Leaderboard's specific CARLA version.
    \item PCLA includes nine different high-performing ADAs trained with 17 distinct training seeds.
\end{itemize}

\section{PCLA}

Various research efforts focus on testing autonomous vehicles, including test scenario generation and testing ADAs' robustness against adversarial attacks~\cite{haq2023many,zhang2023data}. This research often involves generating complex driving scenarios or altering the environment to induce failures in the vehicle's behavior~\cite{tehrani2024taxonomy}. Consequently, researchers in this field typically do not need to train an ADA but instead require a ready-to-use agent to evaluate their test methods. PCLA provides them with an easy-to-use test infrastructure for CARLA. 
In this section, we present an overview of the CARLA Leaderboard, the challenges faced during the deployment of an autonomous agent in CARLA, our approach to building PCLA, and a simple example of its usage.

\subsection{CARLA Leaderboard}

CARLA does not come with a prebuilt ADA, requiring researchers to select and integrate an agent into the simulation -- a challenging task. Leaderboard agents are commonly selected because of their strong performance in solving realistic driving scenarios. However, to use the Leaderboard agents, researchers must thoroughly understand the Leaderboard’s base code and manually modify it to test different agents. Since each ADA requires its specific setup, this process becomes increasingly time-consuming and complex when switching between multiple agents for testing.

The Leaderboard codebase manages all aspects of the simulation, including environment and traffic. It features predefined scenarios with various towns and different numbers of actors (i.e., vehicles and pedestrians). The Leaderboard codebase is responsible for configuring multiple elements of the simulation. It places sensors on the vehicle, translates the given route into a CARLA-compatible route within the specified town, sets up the selected autonomous agent on the vehicle, handles the progress of time, and manages essential CARLA aspects, such as the \textit{world} and \textit{client}. The CARLA \textit{world} is an object that represents the simulation. It serves as an abstraction layer, providing key methods to spawn actors, modify the weather, retrieve the current state of the world, and more. The CARLA \textit{client}, on the other hand, is the module that users run to request information or make changes within the simulation. Each client operates using a specific IP address and port, communicating with the server via the terminal. Multiple clients can run simultaneously, allowing for concurrent interactions with the simulation.

In a typical scenario, the Leaderboard spawns the autonomous vehicle in a specified town at the starting point of a given route. Additional actors are spawned at random locations, and a spectator camera is positioned to provide a bird's eye view of the autonomous vehicle. This camera follows the vehicle throughout its journey.

\subsection{Challenges}

To customize the CARLA world, significant modifications to the Leaderboard codebase are required, which necessitate a deep understanding of the code structure. For instance, users cannot directly control the specific locations where vehicles and pedestrians are spawned; they can only adjust the total number of actors. Additionally, users cannot assign specific missions to non-autonomous vehicles and pedestrians or dictate their routes to make them move in the desired direction.

Most recent research~\cite{giamattei2025reinforcement,tehrani2024taxonomy} emphasizes the importance of leveraging the latest technology. While the most recent version of CARLA is 0.9.15, the Leaderboard continues to rely on CARLA 0.9.10. This incompatibility makes it difficult for researchers to conduct experiments on the latest advancements in autonomous vehicle testing.

\subsection{Approach}
PCLA  is a framework designed for researchers seeking to deploy one or more ADAs onto vehicles in CARLA. PCLA empowers the selected vehicle with the capabilities of the chosen autonomous agent, allowing users to customize the environment, the actors, and the actors' behaviors to their specific needs. PCLA serves as a middleman, providing the \textit{control} action output of a Leaderboard agent directly to the developer.

Under the hood, PCLA leverages the Leaderboard codebase to simplify the vehicle and agent configuration. It configures the vehicle’s sensors, positions them according to the agent’s requirements, and sets the selected agent to receive input from these sensors. To set up the route, PCLA also translates the provided route into CARLA-compatible GPS coordinates by interpolating a dense trajectory and mapping each waypoint along the route.

\begin{figure}[t]
    \centering
    \includegraphics[width=\linewidth]{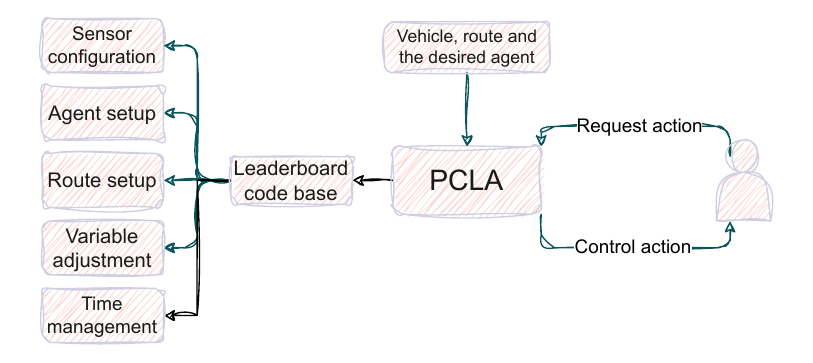}
    \caption{Overview of how the PCLA framework operates} \label{fig:PCLA}
\end{figure}

Additionally, PCLA manages the time constraints required for recording the performance during Leaderboard challenges and adjusts CARLA’s core variables, such as the \textit{world} and \textit{client}, to integrate them into the Leaderboard’s variables. This abstraction ensures that developers are not dependent on the Leaderboard’s codebase. They only need to provide PCLA with the vehicle, route, and desired agent. PCLA then takes care of the rest, enabling developers to simply request an action and receive the corresponding \textit{control} action for the autonomous vehicle from the selected agent. When an action is requested, PCLA forwards the request to the selected agent. The agent computes the \textit{control} action based on the state of the ego vehicle and the sensor outputs at that specific time frame. 
A \textit{control} action contains values representing the vehicle's actions for a single frame, including throttle, steering, brake, hand brake, reverse, manual gear shift, and gear. These are the movement parameters that are essential for the vehicle to determine its actions in the next frame.
The resulting \textit{control} action can then be applied to the vehicle, enabling it to move accordingly or be used in other testing applications.
An abstract representation of how PCLA operates is illustrated in Figure~\ref{fig:PCLA}

\subsection{Usage}
To use our framework,  users  need to import it and provide:
\begin{itemize}
    \item The name of the desired agent
    \item The route for the vehicle to follow
    \item And the vehicle that is aimed to be autonomous
\end{itemize}
Additionally, users need to pass the CARLA's \textit{client} module, which is required for configuring the Leaderboard attributes. Finally, at any desired time, users can call the \textit{get\_action()} method from the PCLA class to retrieve a control action based on a single frame captured by the agent's sensors. For a better understanding, a sample code snippet from the GitHub repository is provided below.
\begin{lstlisting}[language=Python]
from PCLA import PCLA

agent = "neat_neat"
route = "./sampleRoute.xml"
pcla = PCLA(agent, vehicle, route, client)

ego_action = pcla.get_action()
vehicle.apply_control(ego_action)
\end{lstlisting}

\noindent
In the code snippet above, after importing the framework in line 1, the agent is selected by its name in line 3, and the path to the route file is specified in line 4. In line 5, the PCLA class is initialized by passing these variables along with the vehicle and \textit{client}. 
Then, in line 7, the \textit{pcla.get\_action} method is called to retrieve a control action from the ADA. Finally, in line 8, the action is applied to the vehicle using the \textit{apply\_control} command in CARLA.

\section{Research Scenarios}
PCLA is potentially useful in various research scenarios such as testing autonomous driving systems against complex driving scenarios or testing their robustness against adversarial attacks.

Testing deep neural network-enabled systems (such as ADAs) is a challenging and costly endeavor, yet it remains crucial in the development of many modern systems with artificial intelligence at their core~\cite{giamattei2025reinforcement}.
Test generation is typically performed automatically by solving an optimization problem~\cite{haq2023many}. This involves identifying the optimal configurations of objects in the environment to maximize or minimize an objective function. The testing process must select an appropriate algorithm to learn the actions that maximize reward, particularly by exposing safety requirement violations within a given time frame.

Test cases in CARLA can simulate a variety of scenarios that an autonomous vehicle might encounter. Examples can include situations where another vehicle suddenly brakes in front~\cite{lu2022learning}, a pedestrian unexpectedly steps into the road~\cite{koren2018adaptive}, or a chaotic scene happens at an intersection~\cite{corso2019adaptive}.

Another type of autonomous vehicle testing focuses on evaluating the robustness of an ADA against adversarial attacks or vice versa. In such research, authors might create adversarial images or objects and place them in the environment using the CARLA editor to induce system failures in autonomous vehicles~\cite{tehrani2024taxonomy}. These adversarial images or objects, which are invisible to normal human vision, are strategically placed in locations such as roads, billboards, sidewalks, or lane markings~\cite{von2023deepmaneuver,sato2021dirty,boloor2020attacking,pavlitskaya2020feasibility}. %

Researchers may require the current movement parameters of the autonomous vehicle to conduct specific analyses. For example, in  DeepManeuver~\cite{von2023deepmaneuver}, the attack algorithm needs detailed information, such as the steering angle and speed, to compute the next adversarial patch on a billboard. PCLA facilitates such use cases by providing the \textit{control} action, which includes complete details about the vehicle's movement parameters.

With PCLA, researchers can create customized environments, vehicles, and actors in the latest CARLA version and perform quick switching between different agents, allowing researchers to optimize their test scenarios based on various ADAs or observe how various ADAs respond to the same adversarial attack. This flexibility greatly enhances the research process of generating test cases in CARLA and testing the robustness of autonomous systems against adversarial attacks or vice versa.

\section{Related Work}

This section is divided into two subsections. The first provides a brief introduction to the PCLA agents and summarizes the corresponding research papers. The second focuses on papers that explore testing agents within the CARLA simulator.

\subsection{Available Agents}
Currently, PCLA includes nine Leaderboard agents and 17 distinct training seeds. These agents were selected for their high performance in the CARLA Leaderboard challenge and their compatibility with newer versions of CARLA and Python packages. The list of related papers is as follows:

\begin{itemize}
    \item \textbf{Hidden Biases of End-to-End Driving Models}\footnote{\url{https://github.com/autonomousvision/carla_garage}}. Jaeger et al.~\cite{jaeger2023hidden} developed an open-source Leaderboard 2.0 starter kit called CARLA-garage, which includes a dataset, expert driver, evaluation tools, and training code. Additionally, they provided pre-trained model weights for \textbf{TransFuser++}, the best open-source model available at the time of their publication. This repository contained four pre-trained models and their weights, and 12 training seeds, all of which are included in the PCLA framework.
    \item \textbf{NEAT: Neural Attention Fields for End-to-End Autonomous Driving}\footnote{\url{https://github.com/autonomousvision/neat}}, proposed by Chitta et al.~\cite{Chitta2021ICCV}, includes four different agents. The paper argues that efficient reasoning about the semantic, spatial, and temporal structure of a scene is a critical prerequisite for autonomous driving. It introduces a continuous function that maps locations in Bird’s Eye View (BEV) scene coordinates to waypoints and semantics. This mapping is achieved through intermediate attention maps, which iteratively compress high-dimensional 2D image features into a compact representation. PCLA includes all four agents from this paper, making them easily accessible and switchable.
    \item \textbf{InterFuser: Safety-Enhanced Autonomous Driving Using Interpretable Sensor Fusion, Transformer}\footnote{\url{https://github.com/opendilab/InterFuser}} introduces an agent called \textbf{TransFuser}. Shao et al.~\cite{shao2022interfuser} proposed a safety-enhanced autonomous driving framework named Interpretable Sensor Fusion \textbf{Transformer (InterFuser)}. This framework is designed to fully process and fuse data from multi-modal, multi-view sensors to achieve comprehensive scene understanding and adversarial event detection. This repository, which includes one agent, is integrated into PCLA and allows users to view the agent's sensor feed live.
\end{itemize}
\subsection{Testing on Carla Agents}
Many studies have explored testing autonomous systems within the CARLA simulator environment~\cite{tehrani2024taxonomy}. 

In the area of test generation, Haq et al.~\cite{haq2022efficient} introduced SAMOTA, a novel approach for effectively and efficiently generating test data for DNN-enabled systems in the context of online testing. This method combines surrogate-assisted optimization with many-objective search to evaluate and test advanced DNN-enabled autonomous driving systems within the CARLA simulator.

Later again, Haq et al.~\cite{haq2023many} introduced MORLOT, a novel online testing approach that leverages reinforcement learning (RL) to incrementally generate sequences of environmental changes. This approach utilizes many-objective search to identify changes that are more likely to uncover previously untested scenarios within the CARLA simulator.

For adversarial attacks, in the research conducted by Boloor et al.~\cite{boloor2020attacking}, using the CARLA simulator, adversarial road lines were painted by the attacker at intersections or on curved roads. These lines were designed to point in the opposite direction of the lane's intended turn, causing the victim’s car to turn incorrectly and hit the road walls or lose its path.

Pavlitskaya et al.~\cite{pavlitskaya2020feasibility} proposed an approach to manipulate an autonomous vehicle’s trajectory by placing a printed adversarial patch on the roadside in the CARLA simulator. This intervention caused the vehicle to steer toward the patch, ultimately resulting in a collision.

Piazzesi et al.~\cite{piazzesi2021attack} manipulated the trained agent’s neurons and weights, along with the input images, to induce incorrect steering decisions and traffic light misdetections. These modifications led to severe outcomes, including collisions with surroundings, lane departures, off-road driving, running through intersections, and ignoring traffic lights all in the CARLA simulator.

By using PCLA, the authors of the above-mentioned papers would have been able to design and implement the code for their experiments more quickly and easily, with the possibility to experiment with multiple driving agents operating on the latest version of CARLA, thanks to the abstraction layer provided by our framework.

\section{Conclusion \& Future Work}
ADA testing researchers often choose driving agents from the CARLA Leaderboard, as these agents achieve state-of-the-art performance. 
PCLA simplifies the complex task of finding and using a pre-trained autonomous driving agent from the Leaderboard in the CARLA simulator. With PCLA, researchers
can easily select and switch between nine of the highest-performing CARLA Leaderboard agents, and can easily deploy them onto a vehicle, or use their outputs e.g. for system level attack generation, 
without requiring any prior knowledge of the Leaderboard codebase.

Additionally, PCLA eliminates the restriction of using a specific CARLA version by being fully compatible with the latest version of the simulator. This framework is particularly valuable for testing and evaluating various agents in customized scenarios.

Looking ahead, we plan to expand the PCLA framework by including more Leaderboard agents and introducing more generally a wider variety of existing driving agents, to further enhance its versatility.

\section*{Framework Repository}

The framework, source code, step-by-step tutorials, and a sample code are publicly available at \url{https://github.com/MasoudJTehrani/PCLA}

\begin{acks}
This work is funded by the European Union's Horizon Europe research and innovation programme under the project Sec4AI4Sec, grant agreement No 101120393.
\end{acks}

\bibliographystyle{ACM-Reference-Format}
\bibliography{bib}

\end{document}